# Full Title:

## An Orbitrap-based laser desorption/ablation mass spectrometer designed for spaceflight

### Short Title:

A laser desorption/ablation Orbitrap mass spectrometer for spaceflight

### Authors and Affiliations:

Ricardo Arevalo Jr.[1*], Laura Selliez[2,3], Christelle Briois[2], Nathalie Carrasco[3], Laurent Thirkell[2], Barnabé Cherville[2], Fabrice Colin[2], Bertrand Gaubicher[2], Benjamin Farcy[1], Xiang Li[4], and Alexander Makarov[5]

[1] Department of Geology, University of Maryland, College Park, MD, USA 20742

[2] Laboratoire de Physique et Chimie de l'Environnement et de l'Espace (LPC2E), UMR 7328 du CNRS, 45071 Orléans, FR

[3] Laboratoire Atmosphères, Milieux, Observations Spatiales (LATMOS), 78280 Guyancourt, FR

[4] Center for Space Science & Technology, University of Maryland, Baltimore County, Baltimore, MD, USA 21250

[5] Thermo Fisher Scientific GmbH, 28199 Bremen, DE

### *Corresponding Author:

Ricardo Arevalo Jr.

Department of Geology

Office: CHEM 1217A

University of Maryland

College Park, MD 20742

Phone: (301) 405-5352

Fax: (301) 405-3597

rarevalo@umd.edu

**Compound Abstract:**

**RATIONALE:** The investigation of cryogenic planetary environments as potential harbors for extant life and/or contemporary sites of organic synthesis represents an emerging focal point in planetary exploration. Next generation instruments need to be capable of unambiguously determining elemental and/or molecular stoichiometry via highly accurate mass measurements and the separation of isobaric interferences.

**METHODS:** An Orbitrap$^{TM}$ analyzer adapted for spaceflight (referred to as the *CosmOrbitrap*), coupled with a commercial pulsed UV laser source (266 nm), is shown to successfully characterize a variety of planetary analog samples via ultrahigh resolution laser desorption/ablation mass spectrometry. The materials analyzed in this study include: jarosite (a hydrous sulfate detected on Mars); magnesium sulfate (a potential component of the subsurface ocean on Europa); uracil (a nucleobase of RNA); and a variety of amino acids.

**RESULTS:** The instrument configuration tested here enables: measurement of major elements and organic molecules with ultrahigh mass resolution ($m/\Delta m \geq 120{,}000$, FWHM); quantification of isotopic abundances with $<1.0\%$ ($2\sigma$) precision; and, identification of highly accurate masses within 3.2 ppm of absolute values. The analysis of a residue of a dilute solution of amino acids demonstrates the capacity to detect twelve amino acids in positive ion mode at concentrations as low as $\leq 1$ pmol/mm$^2$ while maintaining mass resolution and accuracy requirements.

**CONCLUSIONS:** The *CosmOrbitrap* mass analyzer is highly sensitive and delivers mass resolution/accuracy unmatched by any instrument sent into orbit or launched into deep space. This prototype instrument, which maps to a spaceflight implementation, represents a mission-enabling technology capable of advancing planetary exploration for decades to come.

# INTRODUCTION

## *Organics in Cryogenic Environments*

Understanding the origin, distribution, and processing history of organic compounds in cryogenic planetary environments is one of the most compelling future directions in solar system research. Such organics are structurally and functionally diverse, despite their low temperature origins, and are thus thought to constitute an enabling prebiotic inventory for the synthesis of more complex macromolecular organics, and ultimately the potential emergence of life. Top priority planetary science goals for the coming decades will require detailed *in situ* studies of surface and subsurface composition to unambiguously identify elemental and molecular constituents of complex, multicomponent mixtures of organics preserved within icy geological matrices. These planned investigations will further our understanding of primordial sources of prebiotic organic compounds, viable sites of progressive organic polymerization, and specific abiotic versus biotic pathways that enable the construction of functional/active macromolecular networks.

Recent missions to asteroids (*e.g.*, Dawn[1,2]), comets (*e.g.*, Rosetta[3,4]), and various ocean worlds in the Jovian system (*e.g.*, Galileo[5,6]) and Saturnian system (*e.g.*, Cassini-Huygens[7,8]), have indicated that these solar system bodies may serve as prospective refuges for primordial organic matter and/or sites of organic synthesis due to: i) the availability of carbon-rich starting materials, including prebiotic organic compounds in many environments; ii) water ice and putative physicochemical interfaces between aqueous reservoirs and silicate systems; and, iii) active sources of energy, such as hydrothermal activity, ultraviolet radiation, electrical discharges, cryovolcanism, and/or impacts. The plausibility of these sites as potential harbors for extinct and/or extant life signatures has been supported by both spaceborne and laboratory investigations.

The *in situ* detection of hydrated minerals, surface morphologies indicative of volatile outgassing, and local exposures of organic materials on the asteroids Vesta[9-12] and Ceres[1,2,13-15], coupled with laboratory analyses of nucleobases[16-20] and non-racemic amino acids[21-23] found in chondritic meteorites, point to asteroids as prospective shelters for prebiotic and macromolecular organic compounds. Samples of comet 81P/Wild 2 returned to Earth via the Stardust mission, and *in situ* analysis of the coma of comet 67P/Churyumov-Gerasimenko, suggest amino acids of extraterrestrial origin may be prevalent on cometary bodies, too[24,25]. Measurements of the Titan atmosphere enabled by the Voyager 1 flyby, Cassini orbiter, and Huygens probe indicate a substantial inventory of organic molecules, including a host of complex hydrocarbons, nitriles, and solid organic aerosols[8,26-29]. Finally, plumes observed jetting from Europa and Enceladus point to subsurface liquid water reservoirs that contain simple hydrocarbons (up to $C_5$) and complex macromolecular organic materials (molecular masses up to > 200 Da), as well as nitrogen and chemical sources of energy[30-34].

Organic matter detected on these planetary bodies may have been derived from one or more of the following mechanisms: the infall of carbon-rich small bodies; abiotic processes such as Fisher-Tropsch reactions or Strecker synthesis; Titan-like photochemical haze incorporation; or geological activity, such as water–rock interactions, thermogenesis, and/or biogeochemistry. Consequently, our understanding of the origin of such organics, including endogenous versus exogenous sourcing and biotic versus abiotic processing, remains incomplete. Fortunately, cryogenic systems represent ideal preservational environments for ancient and/or contemporary macromolecular organics, biominerals, and morphological structures indicative of progressive molecular polymerization and the potential emergence of microbial life. Ocean worlds have thus been identified by NASA[35,36], the Outer Planetary Assessment Group[37], and the US House Appropriations Committee[38] as high priority targets for near-term exploration.

In order to improve our understanding of potentially habitable cryogenic environments, which may serve as sites of progressive organic synthesis and/or sanctuaries for biosignatures reflecting microbial activity (should life ultimately emerge), future missions need to enable comprehensive, *in situ* compositional analysis of surface, subsurface, and plume-derived materials from the various planetary bodies described above. Critical investigations needed in the coming decade include: the search for and unambiguous identification of amino acids, nucleobases, and other prebiotic organic molecules; accurate determinations of elemental and molecular abundance patterns; precise isotopic measurements of C, N, and other elements essential to life; and, (semi)quantitative mineralogy for geological context, including the detection of potential biominerals.

*Mass Spectrometry and Isobaric Interferences*

In laboratory settings, mass spectrometry techniques are routinely applied to assay the organic content of solid, liquid, and gaseous sample materials, particularly in the realms of: proteomics; pharmaceutical medicine; forensic science; structural biology; energy and biofuels; environmental studies; and, astrobiology. However, the unambiguous identification of complex organic molecules requires the differentiation of competing isobaric species characterized by the same nominal mass-to-charge ratio (or *m/z*). In order to isolate potential isobars, thereby allowing for the detection and quantitation of targeted organic signals, commercial instruments offer a number of advanced capabilities and/or hardware upgrades.

Many systems are equipped with a gas or liquid chromatograph that can physically separate the different components of a mixture by passing the analyte through a stationary phase prior to introduction into the mass analyzer. Mass spectrometers that support tandem mass spectrometry, or MS/MS, rely on multiple analytical steps (and sometimes the combination of two analyzers in a single system) to selectively ionize a targeted compound, isolate precursor or "parent" ions, decompose those ions via interactions with incident photons or gas, and finally analyze the product or "daughter" fragments via traditional mass-selective detection. Laser-based resonance ionization mass spectrometers (RIMS), such as those under development for planetary geochronology investigations[39], can selectively ionize specific elements, but require multiple laser systems with highly stable wavelength emissions. Other sensors enable the discrimination of isobaric interferences through high mass resolving powers, defined as the instrument's capacity to distinguish two adjacent mass peaks with only slightly different exact masses ($\Delta m$) but equal intensities; the mass resolving power of an instrument is commonly annotated as $m/\Delta m$ at a stated peak height (*e.g.*, FWHM, or Full Width Half Maximum).

Of these options, high mass resolution is the only capability that can distinguish competing isobaric interferences without depending on additional subsystems (*e.g.*, a gas or liquid chromatograph, or multiple mass analyzers or lasers) and/or multiple stages of analysis. However, the mass resolving power required for any specific application needs to be considered. For example, the volatile gases CO and $N_2$ share a common nominal mass of 28

Da but their exact monoisotopic masses are separated by 11 mDa (or 400 ppm), thus requiring a mass resolving power of $m/\Delta m \geq 2500$ (FWHM) to quantitatively isolate these mass peaks.

The differentiation of overlapping organic signals can be even more challenging, especially at higher masses due to a greater number of possible permutations of the life essential elements CHNOPS and isotopologues. For example, leucine ($C_6H_{13}NO_2$, an α-amino acid used in the biosynthesis of proteins), hydroxyproline ($C_5H_9NO_3$, hydroxylation product of the α-amino acid proline), and creatine ($C_4H_9N_3O_2$, a nitrogenous organic acid synthesized from α-amino acids glycine and arginine) all share a common nominal mass (131 Da), requiring a mass resolving power of $m/\Delta m > 11,000$ to distinguish each compound. At even higher molecular weights, the METLIN metabolite mass spectral database[40] may be used to elucidate common isobaric interferences. For instance, the database has cataloged 19 competing isobars (not including alkali metal adducts or structural isomers) within 20 ppm of mass 245.08 Da, coinciding with the mass of protonated uridine (a ribonucleoside) and requiring $m/\Delta m > 100,000$ to discriminate all compounds. Distinguishing inorganic species, including atomic and molecular signals, requires ultrahigh mass resolution in many cases, too; mass resolving powers of $m/\Delta m > 65,000$ and $m/\Delta m > 74,000$ are required to distinguish moderately volatile $^{70}$Zn (0.6% isotopic abundance) from $^{70}$Ge (20.4%), and more refractory $^{54}$Cr (2.4%) from $^{54}$Fe (5.8%), respectively.

*In situ Mass Spectrometry for Planetary Exploration*

Since the Pioneer Venus Program in the mid-1970's, quadrupole mass spectrometers (QMS) have served as our primary means to explore the compositions of objects from the inner and outer reaches of the solar system, including Venus, the Moon (LADEE), Mars (MSL and MAVEN), the Jovian system (Galileo), and the Saturnian system (Cassini-Huygens). Although sensitive and quantitative instruments, standard QMS sensors detect organic compounds with only limited mass resolution (typically $m/\Delta m < 500$, FWHM[41-47]). Consequently, QMS peak assignments are often tentative, as a single peak may comprise a multitude of organic and/or inorganic isobaric components that contribute to the signal to varying degrees (**Fig. 1**).

In contrast to heritage QMS systems, next generation sensors that offer higher mass resolving powers (*i.e.*, $m/\Delta m > 10^3$, FWHM) may enable the unambiguous identification of molecular stoichiometry via exact mass determinations with limited mass deviations. The development of the ROSINA investigation[48] onboard Rosetta, and the MASPEX instrument[49] on the upcoming Europa Clipper/Multiple Flyby Mission, highlight the recent transition in planetary science towards spectrometers that offer higher mass resolving powers. However, the Orbitrap$^{TM}$ mass analyzer originally developed by Thermo Fisher Scientific (Bremen, DE) for commercial applications[50-52], and later adapted for spaceflight by a consortium of French laboratories and termed *CosmOrbitrap*[53], may hold the most promise for future astrobiology applications due to unparalleled mass resolution (up to $m/\Delta m > 10^6$, FWHM[54]) and highly accurate mass measurements down to sub-ppm levels[55]. The scientific insights realized by Orbitrap-based mass spectrometers have been demonstrated through numerous laboratory studies using commercial instrumentation to characterize planetary materials, including cometary specimens[56], primitive meteorites[17,57], and Titan analog samples[58-62]. Thus, advanced sensor technologies, such as the Orbitrap mass analyzer, represent the future of planetary exploration, most especially in the realm of biosignature detection/identification.

# EXPERIMENTAL

## *Laser-Enabled Mass Spectrometry*

Regardless of the planetary environment, laser-enabled *in situ* methods of chemical analysis offer an ideal way to characterize precious sample specimens with high spatial resolution and specificity, and without requiring contact with the sample (thereby reducing the risk of contamination). Laser desorption and ablation microprocessing techniques also support: i) reproducible (high-precision) and robust (well-characterized) detection of organic and inorganic molecules over a wide range in mass, volatility, and ionization energy; ii) focused measurements of micron-size targets, such as individual mineral phases, ice grains, and/or discrete geological strata captured by a sample core; and, iii) minimal analytical blanks, resulting in low limits of detection (LOD) and limits of quantitation (LOQ)[63-65]. Laser sampling is particularly well suited for planetary exploration, as such techniques consume orders-of-magnitude smaller quantities of sample (*i.e.*, ng) compared to traditional pyrolysis techniques (*i.e.*, mg), including those executed by the Viking[66], Phoenix[67], and MSL[43] missions.

Laser desorption and ablation mass spectrometry has a long and successful history of use in the elemental and molecular analysis of solid samples in laboratory studies. High peak fluences ($>10$ J/cm$^2$) achieved by modern day pulsed lasers systems permit even the most refractory mineral phases to be sampled (or ablated), while lower fluences ($<1$ J/cm$^2$) can, in many cases, liberate (or desorb) and ionize fragile organic compounds without inducing excessive molecular fragmentation. Solid-state laser systems that generate nanosecond ($10^{-9}$ s, or ns) pulses have served as benchmarks for laser microprocessing[68-72] due to their small and economical mechanical footprints, and their ability to reliably generate tens of millions of pulses at high energies without a compromise in beam quality. Although ns pulse widths incur thermal ablation processes, resulting in elemental and/or molecular fractionation due to sample melting/vaporization/recondensation[65,73-76], limited variations in shot-to-shot laser energy and efficient coupling of the incident radiation with the target (via wavelength specificity) enable (semi)quantitative constraints on sample composition via matrix-matched calibration[63,77,78].

A wide range of non-volatile organic compounds (including macromolecular complexes) and most geological materials (silicates, oxides, etc.) offer high absorption efficiencies at the ultraviolet wavelengths generated by frequency-multiplied Nd:YAG laser systems, particularly 266 nm (4.7 eV/photon) and/or 213 nm (5.8 eV/photon). Although Nd:YAG systems are able to generate higher energy beams at a native wavelength of 1064 nm, infrared radiation is poorly absorbed by many minerals, particularly transparent and/or amorphous phases, and lower photon energies (1.2 eV/photon at 1064 nm) limit bond-breaking potential and ionization efficiency.

The first laser-based mass spectrometer launched into space was the LAZMA instrument[79], designed to analyze samples of regolith collected from the Phobos surface; however, the Phobos-Grunt mission failed to escape low Earth orbit due to an unfortunate thruster malfunction. As a result, the Mars Organic Molecule Analyzer (MOMA) investigation onboard the ExoMars rover will pioneer the first laser desorption mass spectrometer on another planet (assuming successful deployment), including a miniaturized solid-state Nd:YAG laser operating at the fourth harmonic (266 nm), in the search for biosignatures and the characterization of habitability potential in the Martian subsurface[80-82]. Future laser-based mass spectrometers currently under development herald other advanced capabilities, such as highly specific organic ionization/detection[83], controlled molecular fragmentation[84], and quantitative trace element measurements[85].

*Analytical Methods*

In this study, we tested the analytical performance of an Orbitrap mass analyzer that has been adapted for *in situ* planetary exploration through the miniaturization and ruggedization of the supporting mechanical structure, and the implementation of custom electronics that map to heritage electrical systems qualified by previous flight projects[53]. This instrument, referred to as the *CosmOrbitrap* prototype, was interfaced to a Quantel (Les Ulis, FR) Brilliant Q-switched Nd:YAG laser source operating at the fourth harmonic (266 nm wavelength, 6 ns pulse width) and a low-pressure sample stage maintained at $10^{-7}$ mbar or below (**Fig. 2**), thereby reproducing the vacuum expected on the surfaces of many ocean worlds, including Europa[86], Enceladus[87], and Ganymede[88]. This integrated platform serves to demonstrate mission-enabling instrument concepts that center on Orbitrap-based mass analyzers and/or ultraviolet laser systems, particularly those that target cryogenic and/or other low-pressure planetary environments that may harbor extinct or extant life signatures. Specifically, this highly capable and versatile *in situ* instrument delivers:

- Noninvasive, spatially resolved (<100 micron scale) chemical mapping of ice residues and geological/mineralogical materials (including crushed powders and solid cores) via pulsed laser desorption/ablation processing at ultraviolet wavelengths;
- (Semi)Quantitative measurements of inorganic elemental composition and trace levels (LOD ≤ pmol/mm$^2$) of organic compounds, including potential biosignatures up to >500 Da;
- Disambiguation/Differentiation of atomic and molecular isobaric interferences via ultrahigh mass resolving powers ($m/\Delta m$ > 100,000, FWHM) and mass accuracy (< 5 ppm); and,
- High-precision (< 1%, 2σ) determinations of isotopic abundances.

In order to test the capabilities of the *CosmOrbitrap* instrument to detect/identify prebiotic organics and derive information regarding the habitability potential of cryogenic planetary environments, we analyzed a suite of Mars and Europa analog samples, including:

i. synthetically-derived jarosite (KFe(III)$_3$(SO$_4$)$_2$(OH)$_6$), a hydrous iron sulfate mineral discovered on the surface of Mars[89] and suggestive of wet and/or acidic conditions[90-92];
ii. pure uracil (C$_4$H$_4$N$_2$O$_2$), one of the four nucleobases in the nucleic acid of RNA;
iii. magnesium sulfate (MgSO$_4$) salt, which has been detected on the surface of Europa[93] and may represent an ocean brine or radiation product, doped with varying amounts of the α-amino acid valine (C$_5$H$_{11}$NO$_2$); and,
iv. a dilute mixture of sixteen amino acids (10 µmol/L each) suspended in a water solution, a sample even more depleted in non-purgeable organic carbon (NPOC) than subglacial ice from Lake Vostok[94], one of the most extreme and desolate of environments on Earth and a type locality Europa analog site[95].

The tested jarosite powder was generated synthetically from ferric sulfate hydrate (Fe$_2$(SO$_4$)$_3$; Item F0638) procured from Sigma-Aldrich (St. Louis, US) and potassium hydroxide (KOH;

Lot 066279) from Fisher Scientific, following the experimental protocols developed by Driscoll and Leinz[96]. The purity of the sample, estimated at >99%, was verified via x-ray diffraction (XRD) analysis conducted at NASA Goddard Space Flight Center (Greenbelt, US). The powdered uracil (≥99.0% purity; Sigma Item U0750) was analyzed neat, without any further processing. For the $MgSO_4$ mixtures, 97% pure $MgSO_4$ salt (Sigma Item 434183) was physically admixed with powdered valine (≥98.0% purity, Sigma Item V0500) at 3.5 wt.% and 0.35 wt.% concentrations in a solution of deionized $H_2O$ (18 MΩ); the samples were sonicated to encourage homogenous distribution and then air dried in a laminar flow hood. The amino acid solution was derived by diluting a $10^{-3}$ M amino acid solution procured from Waters (Milford, US; Item WAT088122) by a factor of 100 in deionized $H_2O$. Thus, the final solution contained 10 μM of the following amino acids (by increasing molecular weight): glycine; alanine; serine; proline; valine; threonine; leucine and isoleucine; aspartic acid; lysine; glutamic acid; methionine; histidine; phenylalanine; arginine; and, cystine (an oxidized dimer of the α-amino acid cysteine).

Solid powders of the jarosite, uracil, and doped $MgSO_4$ samples were pressed with a clean stainless steel spatula onto sample stubs composed of malleable indium, which provided an internal reference to monitor/verify mass accuracy (via $^{113}In$ and $^{115}In$ peaks). The thickness of each sample application was not tightly controlled between 0.1 – 1.0 mm, allowing for a qualitative investigation into ideal depositional thickness (**Fig. 3**). For the 10 μM amino acid mixture, four drops (2.5 μL each) were pipetted onto an aluminum stub over an area 10 mm in diameter (or 80 mm$^2$) and allowed to dry in a clean chemical fume hood.

Prior to the analysis of each sample, the targeted stub (with sample applied) was installed into the sample chamber via an injection airlock, resulting in a transitional spike in pressure in both the sample and analyzer chambers; once the analyzer chamber pumped down to <5 x $10^{-9}$ mbar, the analyses were allowed to proceed. The pulsed laser beam was focused to an elliptical spot, approximately 40 μm (minor axis) by 80 μm (major axis) in dimension due to a 50° incident angle, on the sample surface by way of two alignment mirrors, a plano-convex lens with a focal distance of 30 cm, and an $MgF_2$ window into the sample chamber. The beam size was verified by image analysis of one- and ten-shot laser pits etched onto burn paper. The laser energy was manually ramped, from <50 μJ up to >500 μJ (via increments of 10 – 20 μJ) until the desorption/ablation threshold of the most easily ionized component in the sample was reached, as indicated by one or more mass peaks higher than background noise. Once the desorption/ablation threshold was identified, the laser output energy was more finely controlled, and periodically increased in order to try to amplify the signal and/or ionize other components within the sample matrix. Typical shot-to-shot reproducibility at a constant laser setting was observed to be on the order of 5-10% (2σ). The energy deposited on the target was monitored by an Ophir (Jerusalem, IL) PE9-SH power meter.

Ions generated by the incident laser radiation were directed towards the *CosmOrbitrap* analyzer via a series of electrostatic lenses that served to both focus/collimate the ion beam and limit gas conductance between the sample and analyzer chambers. Ion velocities were manipulated by high-voltage potentials applied to these lenses, but primarily controlled by a bias applied to the sample stub. A camera (640 × 480 pixel resolution CCD) with a telescopic lens external to the vacuum chamber provided active imaging of the sample surface prior to and during analysis, enabling validation of the alignment between the focal point of the laser beam (*i.e.*, the ablation site) and the axis of the ion optical lens stack through an optical viewport.

The potential of the *CosmOrbitrap* center electrode was controlled by a power supply and high-voltage pulser that together deliver the characteristic voltage ramp needed for capture and electrodynamic squeezing of the incoming ion packet; a voltage stability of better than 100 ppm is required to maximize the mass resolving power. Current induced by the ions as they entered into orbit and continued to oscillate along the center electrode was amplified by a preamplifier board acquired from Thermo Fisher Scientific. Subsequent signal processing, including Hanning apodization and Fast Fourier Transform (FFT), was performed by a customized data acquisition system developed by the Alyxan Company (Orsay, FR).

The space charge capacity of the CosmOrbitrap, controlled primarily by the size and spacing of the surface-matching shapes of the central and outer electrodes, can accommodate up to $10^6$ elementary charges[97], which can be produced by a single laser shot above the desorption/ablation threshold of the sample substrate. The intrascan dynamic range of the analyzer has been experimentally shown to exceed $10^4$, given electronic noise limitations[98]. However, by customizing the mass range (*e.g.*, increasing the low mass cutoff to isolate inorganic peaks from higher mass organic signals) and tuning the laser source (*e.g.*, via wavelength selection and/or energy attenuation control), the dynamic range of the instrument can be expanded by orders of magnitude between scans. In this study, we attempted to load the analyzer with some $10^5$ ions in order to: demonstrate the capabilities of the instrument with margin for improvement; and, protect the preamplifier from oversaturating at the most intense peaks (typically $^{27}$Al or $^{115}$In from the sample stubs).

## RESULTS AND DISCUSSION

### *Synthetic Jarosite*

The ablation threshold of pure jarosite ($KFe(III)_3(SO_4)_2(OH)_6$) was observed between 90 – 100 µJ, equating to a fluence of between 0.7 – 1.2 J/cm$^2$ (taking into account uncertainty in the laser spot size), and resulting in an effective transmission on the order of $10^5$ ions (targeted loading conditions) into the *CosmOrbitrap* analyzer. Each laser shot resulted in a characteristic mass spectrum dominated by peaks for $^{56}$Fe > $^{39}$K > $^{32}$S > $^{16}$O; $^{113}$In and $^{115}$In were also observed, reflecting the composition of the sample stub (**Fig. 4**). Multiple calibration techniques were tested, including single point and linear regression corrections (**Table I**); however, the highest accuracy data were derived when standardizing to either $^{54}$Fe or $^{57}$Fe (single point), resulting in typical mass deviations < 2.9 ppm from absolute values (from the Nuclear Data Center at KAERI, Daejeon, KR).

The relative magnitudes of the peaks observed across the spectra reflect a combination of the concentration of each element in the sample, the abundances of the isotopes measured, the ionization energy of each element, and potential fractionation mechanisms at or near the ablation site. At a wavelength of 266 nm, the energy of the incident radiation (4.7 eV/photon) is enough to ionize K (1$^{st}$ ionization energy, or $E_i$: 4.4 eV) with only a single photon, but Fe ($E_i$: 7.9 eV), S ($E_i$: 10.4 eV), and O ($E_i$: 13.6 eV) all require multiphoton absorption for ionization. Because the thickness of the jarosite powder was not tightly controlled, the magnitudes of the $^{113}$In and $^{115}$In peaks varied significantly, with the highest intensities observed during the analysis of thinner applications of sample.

In order to quantify the isotopic abundances of the major elements measured in the jarosite (as well as $^{113}$In and $^{115}$In from the sample stub), abridged 100 ms signal transients were collected, resulting in a higher density of points outlining each mass peak, albeit at the expense of mass resolving power (by a factor of approx. 6×). Quantified abundances of the major and minor isotopes of In, Fe, K, and S were highly reproducible, and insensitive to laser energy (above the ablation threshold), with measured precision as low as ≤1.0% (2σ) without applying a fractionation correction (**Table II**). Isotopes with low natural abundances of ≤0.1% were difficult to measure quantitatively (*i.e.*, signal-to-noise ratio, SNR < 10) due to Poisson counting statistics limited by the ion volume and instrascan dynamic range of the analyzer. More specifically, ion loading with "only" $10^5$ ions limits such low abundance isotopes to ≤100 ions in the analyzer, defining uncertainties of ≥20% (2σ) due to random counting errors alone, and sampling too few ions to be considered representative of the bulk sample. Other than variation in the intensities of the In peaks, the performance of the instrument, including mass resolution, accuracy, sensitivity, and isotopic precision, was determined to be insensitive (from a statistical perspective) to sample thicknesses between 0.1 and 1 mm.

*Pure Uracil*

Similar to the jarosite sample, the pure uracil powder ($C_4H_4N_2O_2$) was pressed onto an indium sample stub, enabling $^{113}$In and $^{115}$In to act as reference points to quantitatively gauge mass accuracy, and/or serve as internal standards for calibration. The uracil analyte was found to desorb at laser energies as low as <50 μJ (approximately <0.5 J/cm$^2$) with the fragment $C_3H_3NO^+$ (representing the loss of –HNCO) at nominal mass 69 being the first peak to rise above the background, somewhat counterintuitively. Higher energies served to increase the signal of the protonated precursor ion at nominal mass 113 ($C_4H_4N_2O_2^+$; monoisotopic mass: 113.0346 Da), which is easily distinguished from $^{113}$In (exact mass: 112.9041 Da) due to > 1100 ppm difference in absolute mass (**Fig. 5**), as well as the $C_3H_3NO^+$ fragment. All peaks observed in the mass spectra were defined by a mass resolution of $m/\Delta m \geq 130,000$ (FWHM), and the protonated molecule was measured to within 0.1 ppm accuracy of the true mass without requiring an internal standard.

*MgSO$_4$ Doped with Valine*

The MgSO$_4$ sample doped with 3.5 wt.% valine (not shown), which was also applied to an indium stub, was found to ablate at laser energies between 130 – 150 μJ (or 1.1 – 1.7 J/cm$^2$). However, at these laser energies the only inorganic peaks derived from the sample that were measured above the noise floor belonged to Mg and S (negligible O was detected above background), and the only organic peak detected was the the $C_4H_{10}N^+$ fragment of valine at nominal mass 72, corresponding to the primary product expected from electron ionization of valine (per the NIST Chemistry WebBook).

Although the matrix of the MgSO$_4$ sample doped with 0.35 wt.% valine was also observed to ablate at laser energies around 150 μJ (~1.5 J/cm$^2$), the valine fragment $C_4H_{10}N^+$ required ≥250 μJ (≥2.5 J/cm$^2$) to rise above the background. Even higher laser energies closer to 400 μJ (4.0 J/cm$^2$) were needed to achieve the ion loading conditions targeted in this study ($10^5$ ions), likely due to an expanded plasma plume with higher electron density at the sample surface, and by extension increased ionization efficiency at higher laser fluences. At such high laser energies, the protonated molecule ($C_5H_{12}NO_2^+$) at nominal mass 118 was also measured quantitatively (**Fig. 6**). The signal-to-noise ratios of the organic peaks, including both the primary fragment and protonated precursor ion, could have been further improved by increasing the laser energy to even higher values, and/or tuning the targeted mass range to exclude the inorganic matrix ions that dominated the signal (*i.e.*, Mg and S), thereby liberating more ion volume within the *CosmOrbitrap* mass analyzer. All peaks measured, including the highest mass belonging to protonated valine, were measured with a mass resolution of $m/\Delta m \geq$ 120,000 (FWHM) and an accuracy within 3.2 ppm of absolute values.

*1 pmol/mm² Amino Acid Residue*

As described above, four drops (2.5 μL each) of the 10 μM amino acid mixture were pipetted onto an aluminum stub over an area 10 mm in diameter (80 mm²) and allowed to dry. Thus, the residue of the plated solution resulted in approximately 1 pmol/mm² of each amino acid. Of the sixteen amino acids contained in the sample solution, twelve were detected above the background (SNR > 3; **Fig. 7**). Eleven amino acids were observed as protonated molecules, but glycine was only identified as an Al adduct; other amino acids also formed adducts with Al, but at much lower signal levels than the protonated molecules. The amino acids that were not observed above background included serine (nominal mass 105), aspartic acid (mass 133), glutamic acid (mass 147), and the cysteine dimer, cystine (mass 240). The failed detection of these species reflects their preference to form negative ions in laser-induced plasmas with high electron densities and limited proton availability (*i.e.*, little to no matrix to serve as a proton donor)[99-102]; this is not unexpected because amino acids with acidic side chains at neutral pH, such as aspartic and glutamic acids, deprotonate easily and should therefore be expected to form anions more readily. Only positively charged ions were measured in this study, and further work is needed to compare the detection efficiencies of amino acid anions. For the amino acids detected, mass resolution was influenced by both nominal mass and SNR (**Fig. 8**). Of the twelve peaks detected, nine were measured with a mass resolution of $m/\Delta m \geq$ 110,000 (FWHM), and the other three with $m/\Delta m \approx$ 90,000 (FWHM), possibly reflecting poor tuning of the deflector electrode or accelerated decay rates of time-domain transient signals due to larger collision cross-sections[103]. All peaks detected were determined to within 2.0 ppm of their true masses, using the $^{27}$Al peak from the sample stub as an internal standard.

**CONCLUSIONS**

Laser desorption/ablation mass spectrometry, as realized by the *CosmOrbitrap* mass analyzer adapted for spaceflight and a commercial pulsed UV laser system (266 nm), has been shown to effectively characterize: the inorganic elemental composition of geological samples representing analogs of potentially habitable planetary environments, such as formerly acidic and/or aqueous systems on Mars (*e.g.*, jarosite) and the subsurface ocean of Europa (*e.g.*, $MgSO_4$ salt); and, the organic content of synthetically-doped equivalents to salt-rich and salt-poor water residues. In positive ion mode, this instrument configuration was able to detect twelve α-amino acids down to pmol/mm² concentrations. The mass spectra collected on these planetary analog samples illustrate the capacity of the instrument to achieve ≤ 3.2 ppm accuracy and mass resolving powers well in excess of $m/\Delta m > 100{,}000$ (FWHM), and determine isotopic abundances to <1% (2σ) precision. Further work remains to demonstrate the capabilities of the system to analyze negatively charged ions. The prototype laser-interfaced *CosmOrbitrap* developed and tested here represents a game-changing technology capable of revolutionizing *in situ* mass spectrometry, and advancing the agenda of the planetary science community for decades to come.

**Acknowledgements:** We would like to thank James Lewis and Jamie Elsila for assistance in the preparation of the planetary analog samples analyzed here, specifically the lab-created jarosite, physically admixed $MgSO_4$ and L-valine samples, and diluted amino acid solution. We would like to thank the CNES (Centre National d'Etudes Spatiales, France) for the funding of instrumental development of the CosmOrbitrap and the funding with the Région Centre Val de Loire (France) of Laura Selliez, PhD, and Barnabé Cherville, Student Intern. And we thank also the CosmOrbitrap French consortium (LPC2E, LATMOS, IPAG, LISA and CSNSM). Nathalie Carrasco thanks the European Research Council via the ERC PrimChem project (grant agreement No. 636829).

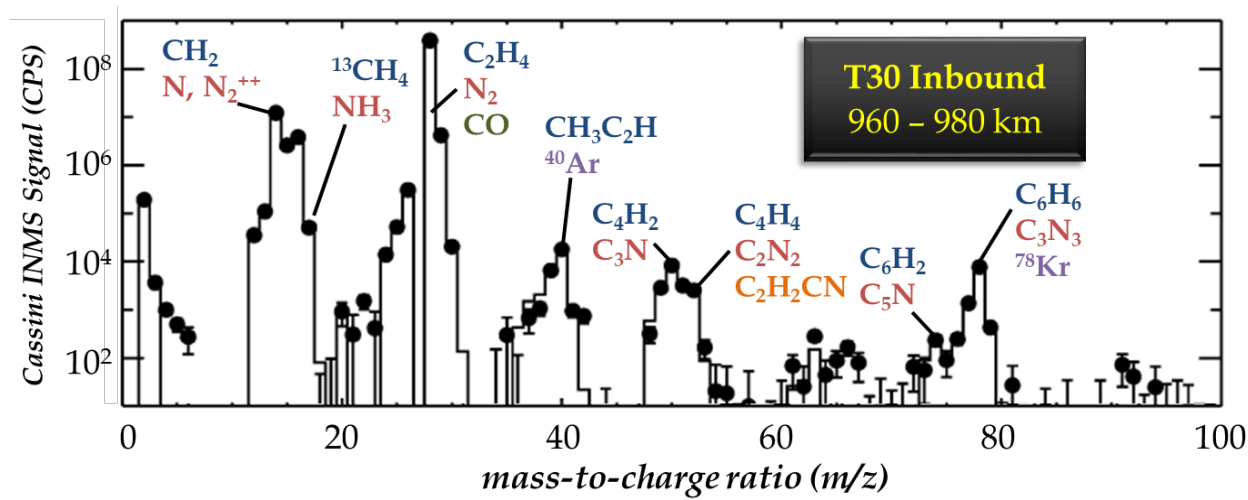

**Fig. 1.** Like most quadrupole mass spectrometers, the Cassini Ion Neutral Mass Spectrometer (INMS) only provides unit mass resolution ($m/\Delta m < 500$, FWHM), leading to uncertain peak assignments. The labels above selected mass stations identify potential isobaric interferences that may be captured underneath each peak; blue species represent hydrocarbons, red species represent nitrogen-bearing compounds, dark green species represent oxides, purples species represent noble gases, and orange species represent complex organics. Spectrum from the Cassini inbound T30 flyby in Titan's ionosphere (between 960 and 980 km) modified from Cui et al.[104].

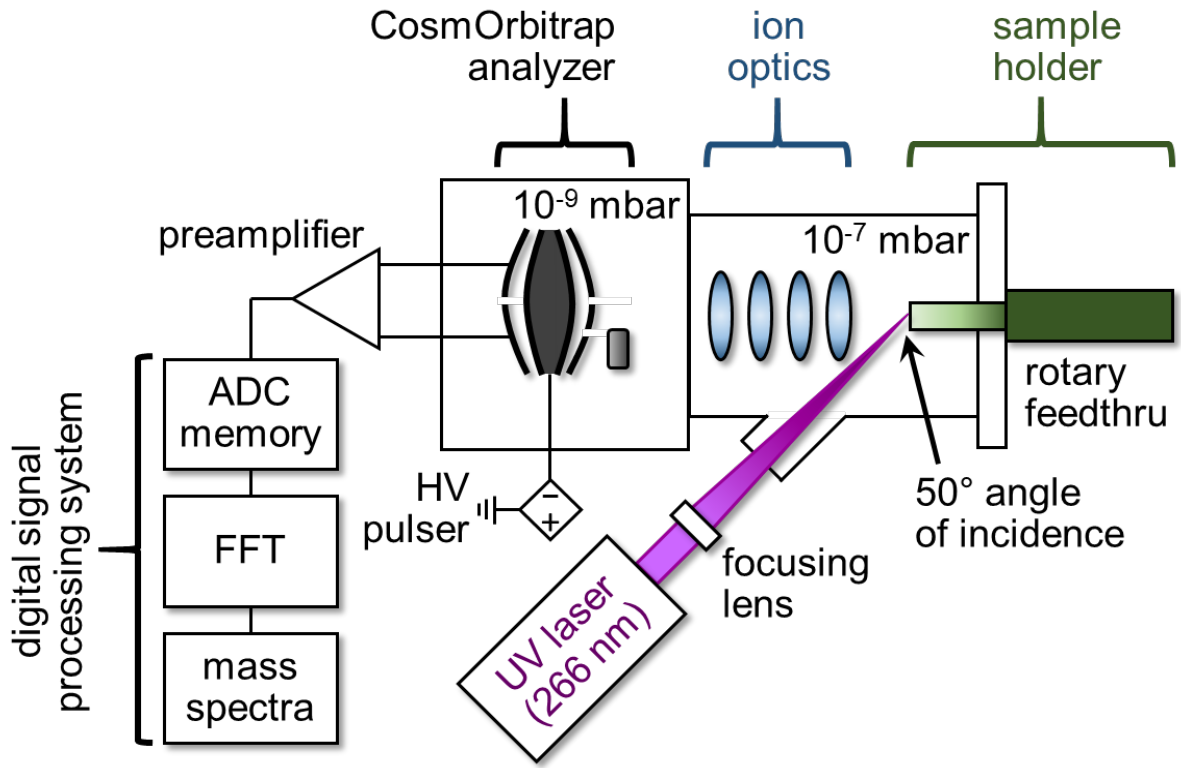

**Fig. 2.** Schematic diagram of the *CosmOrbitrap* prototype instrument (including the analyzer, preamplifier, HV pulser, and digital signal processing system) and planetary simulation chamber maintained in Orléans, France. For more details on this laboratory setup, see Briois et al.[53] .

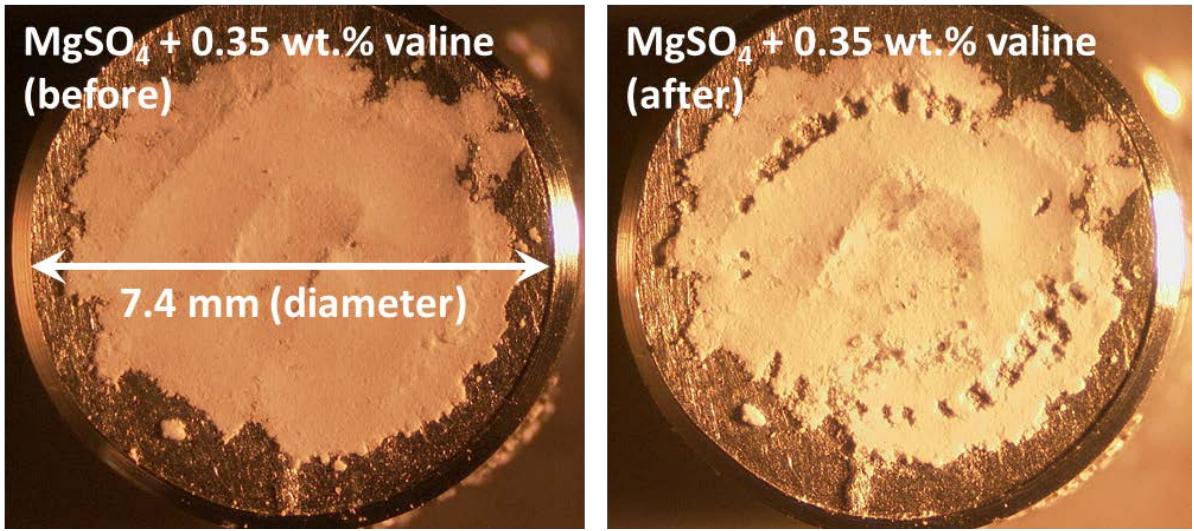

**Fig. 3.** Powder of MgSO₄ doped with 0.35 wt.% valine before (left) and after (right) laser ablation microprocessing and chemical analysis with the *CosmOrbitrap* mass analyzer. The thickness of the sample substrate was not tightly controlled.

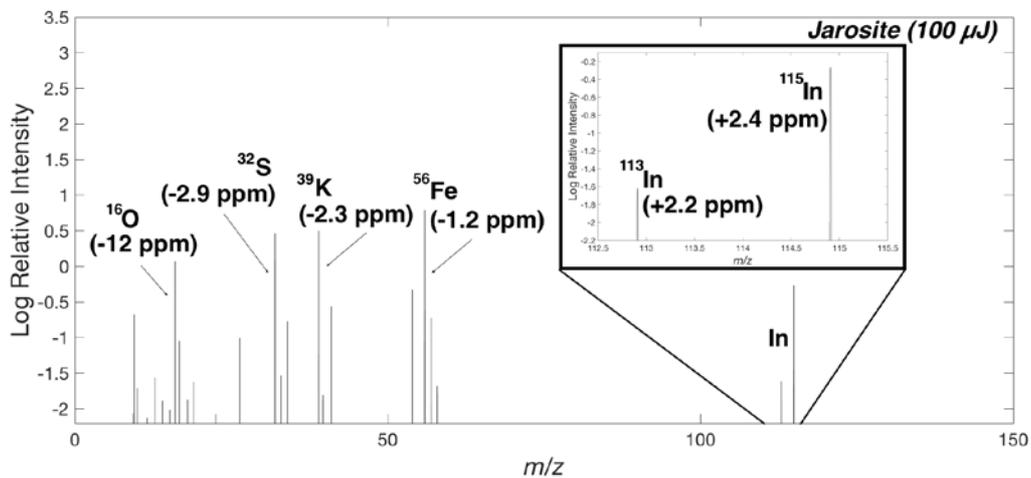

**Fig. 4.** Mass spectrum (100 ms transient) of synthetic jarosite generated by a 100 μJ (approximately 1 J/cm$^2$) ultraviolet laser pulse (266 nm), resulting in the intended ion loading conditions (*i.e.*, 10$^5$ ions in the *CosmOrbitrap* analyzer). Using $^{54}$Fe as an internal standard for calibration, the mass accuracy of nearly all peaks (with the notable exception of $^{16}$O) fall within 2.9 ppm of true values.

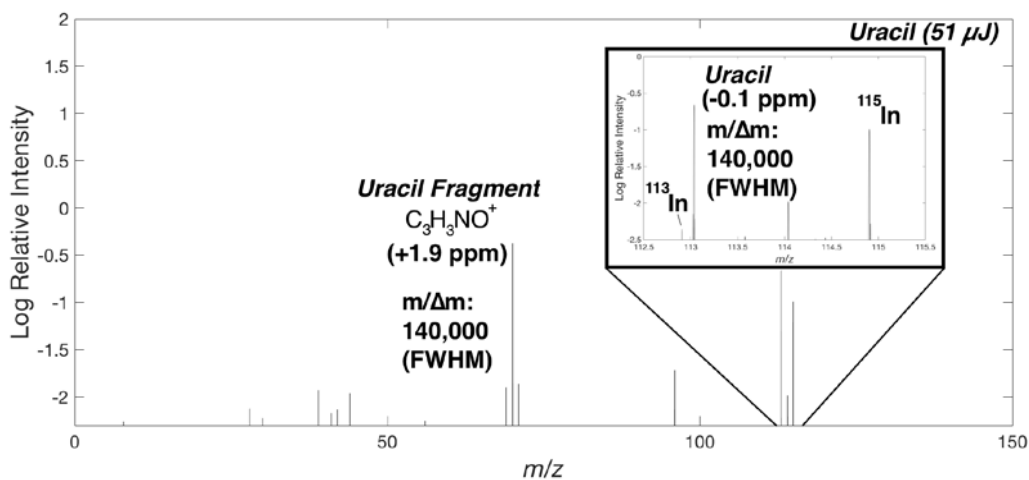

**Fig. 5.** Mass spectrum of pure uracil powder collected at the ablation threshold of the material (51 µJ, or ~0.5 J/cm$^2$). The protonated molecule ($C_4H_4N_2O_2^+$ at *m/z* 113; *m/Δm* = 130,000, FWHM) and $C_3H_3NO^+$ fragment ion (at *m/z* 69; *m/Δm* = 140,000, FWHM), which are both measured with ppm-level mass accuracy, together provide a characteristic signature of the nucleobase. Note, the protonated molecule has the same nominal mass as $^{113}$In, but is easily distinguished at such a high mass resolving power.

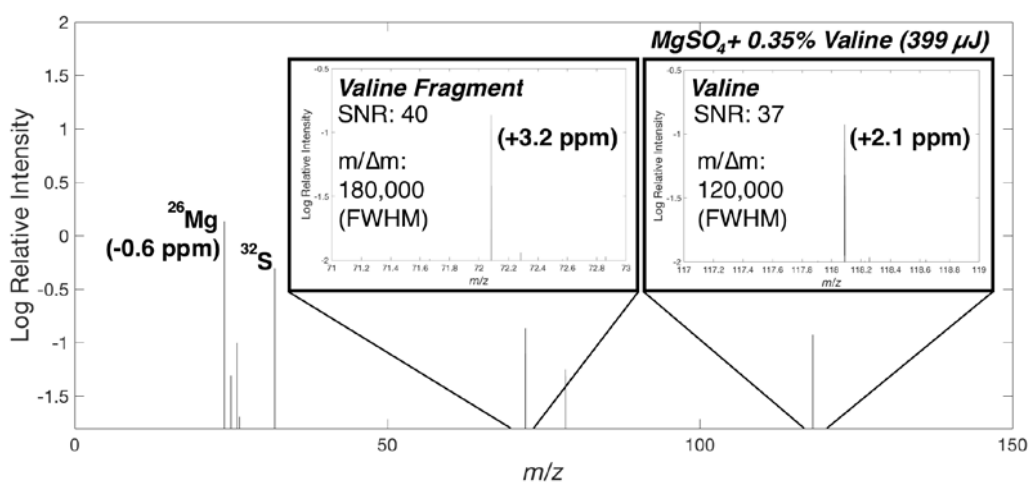

**Fig 6.** Mass spectrum of 0.35 wt.% valine physically admixed with MgSO$_4$ salt. Elemental peaks derived from the magnesium sulfate matrix (*e.g.*, $^{24}$Mg and $^{32}$S), and peaks representing the protonated molecule and fragment ion of valine at *m/z* 118 and 72, respectively, are observed, enabling diagnostic identification of both inorganic and organic components. All peaks were measured with a mass resolution of *m/Δm* ≥ 120,000 (FWHM), and a mass accuracy within 3.2 ppm of true values using $^{25}$Mg as an internal standard for calibration.

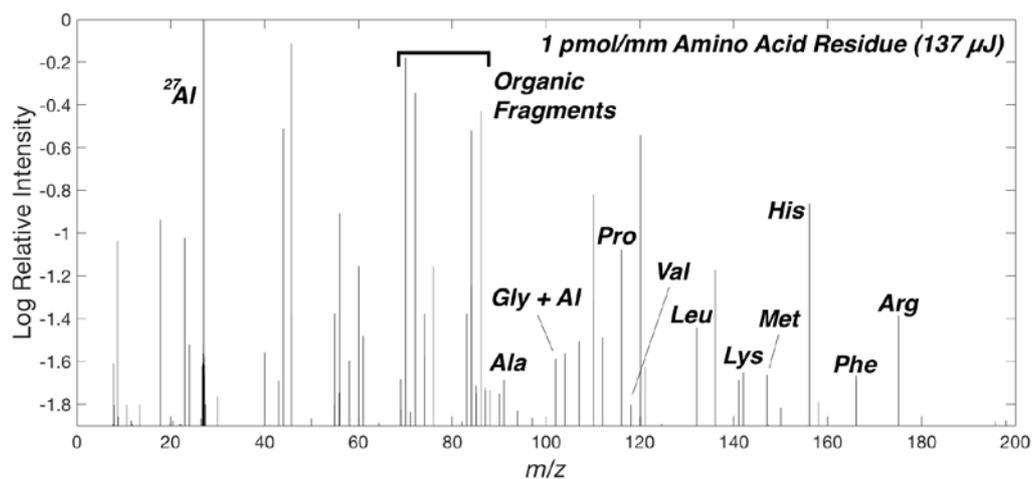

**Fig. 7.** Mass spectrum of a dried residue of 10 μL of a 10 μM amino acid solution plated onto an Al sample stub over an area of 80 mm$^2$, resulting in a surface concentration of approximately 1 pmol/mm$^2$. Of the sixteen amino acids contained in the solution, twelve were detected, eleven as protonated molecules and one as an Al adduct, with a signal-to-noise ratio (SNR) > 3 in positive ion mode (with no distinction between isomers leucine and isoleucine).

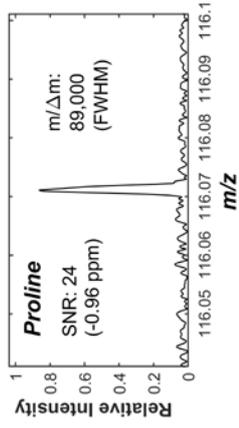
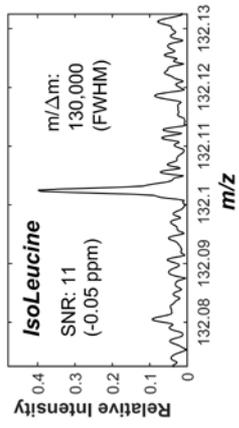
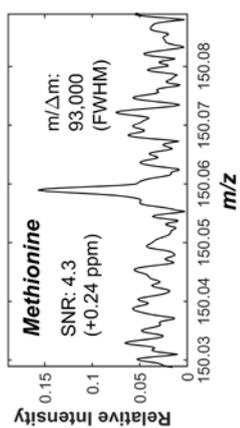
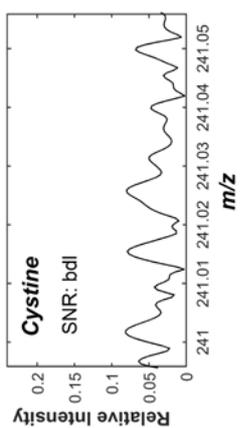
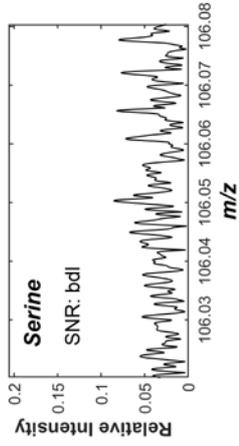
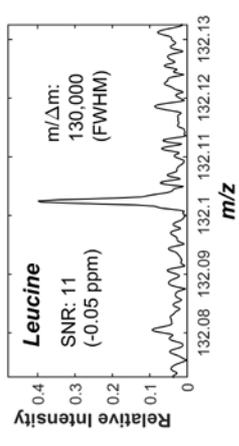
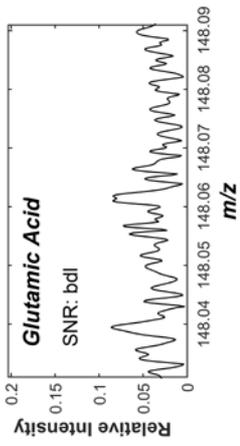
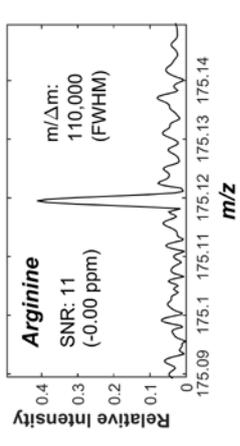
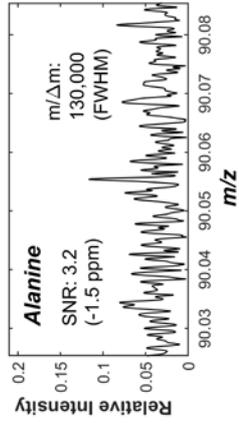
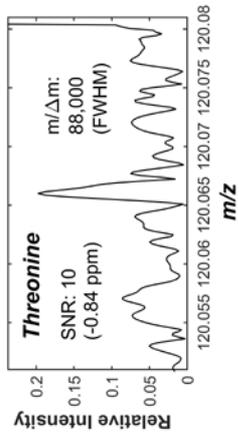
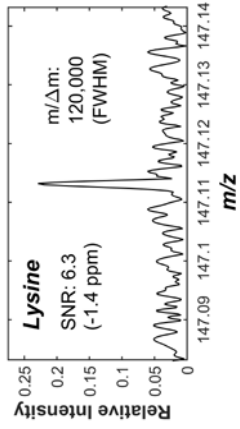
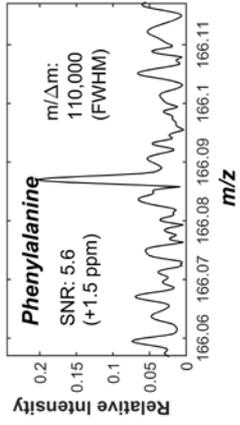
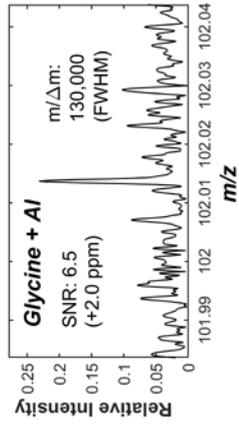
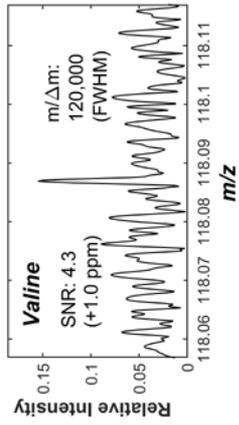
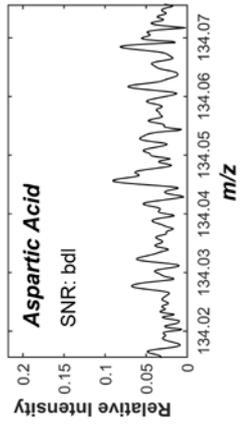
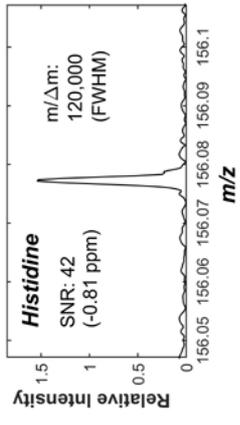

Fig. 8. Zoomed in perspective of the peaks identified in the 10 μM amino acid residue (at a surface concentration of 1 pmol/mm$^2$), as well as four peaks that were not identified above the noise floor (specifically, serine, aspartic acid, glutamic acid, and cystine). Nine of the twelve peaks detected (SNR > 3) are defined by a mass resolution *m/Δm* ≥ 110,000 (FWHM), with the other three measured at *m/Δm* ≈ 90,000 (FWHM). Using $^{27}$Al as an internal standard for mass calibration, all masses were determined with 2.0 ppm accuracy. Unlike the other amino acids detected here, the peak for protonated glycine was not observed; rather, glycine was only identified as an Al adduct. bdl: below detection limit (*i.e.*, SNR < 3).

*References*

1. De Sanctis MC, Ammannito E, McSween HY, et al. Localized aliphatic organic material on the surface of Ceres. *Science.* 2017;355(6326):719-722.
2. Prettyman TH, Yamashita N, Toplis MJ, et al. Extensive water ice within Ceres' aqueously altered regolith: Evidence from nuclear spectroscopy. *Science.* 2017;355(6320):55-59.
3. Capaccioni F, Coradini A, Filacchione G, et al. The organic-rich surface of comet 67P/Churyumov-Gerasimenko as seen by VIRTIS/Rosetta. *Science.* 2015;347(6220).
4. Fray N, Bardyn A, Cottin H, et al. High-molecular-weight organic matter in the particles of comet 67P/Churyumov–Gerasimenko. *Nature.* 2016;538:72.
5. Carlson R, Smythe W, Baines K, et al. Near-Infrared Spectroscopy and Spectral Mapping of Jupiter and the Galilean Satellites: Results from Galileo's Initial Orbit. *Science.* 1996;274(5286):385-388.
6. McCord TB, Carlson RW, Smythe WD, et al. Organics and Other Molecules in the Surfaces of Callisto and Ganymede. *Science.* 1997;278(5336):271-275.
7. Niemann HB, Atreya SK, Bauer SJ, et al. The abundances of constituents of Titan's atmosphere from the GCMS instrument on the Huygens probe. *Nature.* 2005;438:779.
8. Waite JH, Young DT, Cravens TE, et al. The Process of Tholin Formation in Titan's Upper Atmosphere. *Science.* 2007;316(5826):870-875.
9. Denevi BW, Blewett DT, Buczkowski DL, et al. Pitted Terrain on Vesta and Implications for the Presence of Volatiles. *Science.* 2012.
10. McCord TB, Li JY, Combe JP, et al. Dark material on Vesta from the infall of carbonaceous volatile-rich material. *Nature.* 2012;491:83.
11. Prettyman TH, Mittlefehldt DW, Yamashita N, et al. Elemental Mapping by Dawn Reveals Exogenic H in Vesta's Regolith. *Science.* 2012;338(6104):242-246.
12. De Sanctis MC, Combe J-P, Ammannito E, et al. Detection of Widespread Hydrated Materials on Vesta by the VIR Imaging Spectrometer on board the Dawn Mission. *The Astrophysical Journal Letters.* 2012;758(2):L36.
13. Ammannito E, DeSanctis MC, Ciarniello M, et al. Distribution of phyllosilicates on the surface of Ceres. *Science.* 2016;353(6303).
14. De Sanctis MC, Ammannito E, Raponi A, et al. Ammoniated phyllosilicates with a likely outer Solar System origin on (1) Ceres. *Nature.* 2015;528:241.
15. De Sanctis MC, Raponi A, Ammannito E, et al. Bright carbonate deposits as evidence of aqueous alteration on (1) Ceres. *Nature.* 2016;536:54.
16. Burton AS, Stern JC, Elsila JE, Glavin DP, Dworkin JP. Understanding prebiotic chemistry through the analysis of extraterrestrial amino acids and nucleobases in meteorites. *Chem Soc Rev.* 2012;41(16):5459-5472.
17. Callahan MP, Gerakines PA, Martin MG, Peeters Z, Hudson RL. Irradiated benzene ice provides clues to meteoritic organic chemistry. *Icarus.* 2013;226(2):1201-1209.
18. Callahan MP, Smith KE, Cleaves HJ, et al. Carbonaceous meteorites contain a wide range of extraterrestrial nucleobases. *Proceedings of the National Academy of Sciences.* 2011;108(34):13995-13998.
19. Glavin DP, Callahan MP, Dworkin JP, Elsila JE. The effects of parent body processes on amino acids in carbonaceous chondrites. *Meteoritics & Planetary Science.* 2010;45(12):1948-1972.
20. Martins Z, Botta O, Fogel ML, et al. Extraterrestrial nucleobases in the Murchison meteorite. *Earth and Planetary Science Letters.* 2008;270(1):130-136.
21. Cronin JR, Pizzarello S. Enantiomeric Excesses in Meteoritic Amino Acids. *Science.* 1997;275(5302):951-955.

22. Engel MH, Macko SA. Isotopic evidence for extraterrestrial non- racemic amino acids in the Murchison meteorite. *Nature.* 1997;389:265.
23. Kvenvolden K, Lawless J, Pering K, et al. Evidence for Extraterrestrial Amino-acids and Hydrocarbons in the Murchison Meteorite. *Nature.* 1970;228:923.
24. Altwegg K, Balsiger H, Bar-Nun A, et al. Prebiotic chemicals—amino acid and phosphorus—in the coma of comet 67P/Churyumov-Gerasimenko. *Science Advances.* 2016;2(5).
25. Elsila JE, Glavin DP, Dworkin JP. Cometary glycine detected in samples returned by Stardust. *Meteoritics & Planetary Science.* 2009;44(9):1323-1330.
26. Hanel R, CONRATH B, FLASAR FM, et al. Infrared Observations of the Saturnian System from Voyager 1. *Science.* 1981;212(4491):192-200.
27. Maguire WC, Hanel RA, Jennings DE, Kunde VG, Samuelson RE. $C_3H_8$ and $C_3H_4$ in Titan's atmosphere. *Nature.* 1981;292:683.
28. McKay CP. Elemental composition, solubility, and optical properties of Titan's organic haze. *Planetary and Space Science.* 1996;44(8):741-747.
29. Waite JH, Niemann H, Yelle RV, et al. Ion Neutral Mass Spectrometer Results from the First Flyby of Titan. *Science.* 2005;308(5724):982-986.
30. Chyba CF. Energy for microbial life on Europa. *Nature.* 2000;403:381.
31. McKay CP, Anbar AD, Porco C, Tsou P. Follow the Plume: The Habitability of Enceladus. *Astrobiology.* 2014;14(4):352-355.
32. McKay CP, Porco CC, Altheide T, Davis WL, Kral TA. The Possible Origin and Persistence of Life on Enceladus and Detection of Biomarkers in the Plume. *Astrobiology.* 2008;8(5):909-919.
33. Postberg F, Khawaja N, Abel B, et al. Macromolecular organic compounds from the depths of Enceladus. *Nature.* 2018;558(7711):564-568.
34. Roth L, Saur J, Retherford KD, et al. Transient Water Vapor at Europa's South Pole. *Science.* 2014;343(6167):171-174.
35. Directorate NSaM. Science Plan. In. Washington, DC: National Aeronautics and Space Administration; 2014:170.
36. Board CotPSDSSS. Vision and Voyages for Planetary Science in the Decade 2013-2022. In: Sciences NRCotNAo, ed. Washington, DC: National Academies Press; 2011:pp. 410.
37. Hendrix AaH, T. (Chairs). *Roadmaps to Ocean Worlds: Priorities, Mission Scenarios and Technologies* 2018.
38. (2015-2016) tC. H.R.2029 - Consolidated Appropriations Act, 2016. In. Washington, DC: House Appropriations Committee; 2015.
39. Anderson SF, Levine JL, Whitaker TJ. Dating the Martian meteorite Zagami by the 87Rb-87Sr isochron method with a prototype in situ resonance ionization mass spectrometer. *Rapid Communications in Mass Spectrometry.* 2015;29(2):191-204.
40. Smith CA, O'Maille G, Want EJ, et al. METLIN: A metabolite mass spectral database. *Ther Drug Monit.* 2005;27:747-751.
41. Mahaffy PR, Benna M, King T, et al. The Neutral Gas and Ion Mass Spectrometer on the Mars Atmosphere and Volatile Evolution Mission. *Space Science Reviews.* 2015;195(1):49-73.
42. Mahaffy PR, Richard Hodges R, Benna M, et al. The Neutral Mass Spectrometer on the Lunar Atmosphere and Dust Environment Explorer Mission. *Space Science Reviews.* 2014;185(1):27-61.
43. Mahaffy PR, Webster CR, Cabane M, et al. The Sample Analysis at Mars Investigation and Instrument Suite. *Space Science Reviews.* 2012;170(1):401-478.

44. Niemann HB, Atreya SK, Bauer SJ, et al. The Gas Chromatograph Mass Spectrometer for the Huygens Probe. In: Russell CT, ed. *The Cassini-Huygens Mission: Overview, Objectives and Huygens Instrumentarium Volume 1.* Dordrecht: Springer Netherlands; 2003:553-591.
45. Niemann HB, Atreya SK, Carignan GR, et al. The composition of the Jovian atmosphere as determined by the Galileo probe mass spectrometer. *Journal of Geophysical Research: Planets.* 1998;103(E10):22831-22845.
46. Niemann HB, Kasprzak WT, Hedin AE, Hunten DM, Spencer NW. Mass spectrometric measurements of the neutral gas composition of the thermosphere and exosphere of Venus. *Journal of Geophysical Research: Space Physics.* 1980;85(A13):7817-7827.
47. Waite JH, Lewis WS, Kasprzak WT, et al. The Cassini Ion and Neutral Mass Spectrometer (INMS) Investigation. In: Russell CT, ed. *The Cassini-Huygens Mission: Orbiter In Situ Investigations Volume 2.* Dordrecht: Springer Netherlands; 2004:113-231.
48. Balsiger H, Altwegg K, Bochsler P, et al. Rosina – Rosetta Orbiter Spectrometer for Ion and Neutral Analysis. *Space Science Reviews.* 2007;128(1):745-801.
49. Brockwell TG, Meech KJ, Pickens K, et al. The mass spectrometer for planetary exploration (MASPEX). Paper presented at: 2016 IEEE Aerospace Conference; 5-12 March 2016, 2016.
50. Hu Q, Noll RJ, Li H, Makarov A, Hardman M, Graham Cooks R. The Orbitrap: a new mass spectrometer. *Journal of Mass Spectrometry.* 2005;40(4):430-443.
51. Makarov A. Electrostatic Axially Harmonic Orbital Trapping: A High-Performance Technique of Mass Analysis. *Analytical Chemistry.* 2000;72(6):1156-1162.
52. Makarov AA, Inventor; HD Technologies Limited, Manchester, United Kingdom, assignee. Mass spectrometer (US Patent 5,886,346). 1999.
53. Briois C, Thissen R, Thirkell L, et al. Orbitrap mass analyser for in situ characterisation of planetary environments: Performance evaluation of a laboratory prototype. *Planetary and Space Science.* 2016;131:33-45.
54. Denisov E, Damoc E, Lange O, Makarov A. Orbitrap mass spectrometry with resolving powers above 1,000,000. *International Journal of Mass Spectrometry.* 2012;325-327:80-85.
55. Olsen JV, de Godoy LMF, Li G, et al. Parts per Million Mass Accuracy on an Orbitrap Mass Spectrometer via Lock Mass Injection into a C-trap. *Molecular & Cellular Proteomics.* 2005;4(12):2010-2021.
56. Danger G, Orthous-Daunay FR, de Marcellus P, et al. Characterization of laboratory analogs of interstellar/cometary organic residues using very high resolution mass spectrometry. *Geochimica et Cosmochimica Acta.* 2013;118:184-201.
57. Somogyi Á, Thissen R, Orthous-Daunay F-R, Vuitton V. The Role of Ultrahigh Resolution Fourier Transform Mass Spectrometry (FT-MS) in Astrobiology-Related Research: Analysis of Meteorites and Tholins. *Int J Mol Sci.* 2016;17(4):439-.
58. Gautier T, Carrasco N, Schmitz-Afonso I, et al. Nitrogen incorporation in Titan's tholins inferred by high resolution orbitrap mass spectrometry and gas chromatography–mass spectrometry. *Earth and Planetary Science Letters.* 2014;404:33-42.
59. Gautier T, Schmitz-Afonso I, Touboul D, Szopa C, Buch A, Carrasco N. Development of HPLC-Orbitrap method for identification of N-bearing molecules in complex organic material relevant to planetary environments. *Icarus.* 2016;275:259-266.
60. Hörst SM, Yelle RV, Buch A, et al. Formation of Amino Acids and Nucleotide Bases in a Titan Atmosphere Simulation Experiment. *Astrobiology.* 2012;12(9):809-817.

61. Pernot P, Carrasco N, Thissen R, Schmitz-Afonso I. Tholinomics—Chemical Analysis of Nitrogen-Rich Polymers. *Analytical Chemistry.* 2010;82(4):1371-1380.
62. Somogyi Á, Smith MA, Vuitton V, Thissen R, Komáromi I. Chemical ionization in the atmosphere? A model study on negatively charged "exotic" ions generated from Titan's tholins by ultrahigh resolution MS and MS/MS. *International Journal of Mass Spectrometry.* 2012;316-318:157-163.
63. Arevalo Jr. R. Laser Ablation ICP-MS and Laser Fluorination GS-MS. In: Holland) ETa, ed. *Treatise on Geochemistry (2nd Edition).* Vol Vol. 15. Amsterdam, NL: Elsevier Ltd.; 2014:425-441.
64. Russo RE, Mao X, Gonzalez JJ, Zorba V, Yoo J. Laser Ablation in Analytical Chemistry. *Analytical Chemistry.* 2013;85(13):6162-6177.
65. Russo RE, Mao X, Liu H, Gonzalez J, Mao SS. Laser ablation in analytical chemistry—a review. *Talanta.* 2002;57(3):425-451.
66. Anderson DM, Biemann K, Orgel LE, et al. Mass spectrometric analysis of organic compounds, water and volatile constituents in the atmosphere and surface of Mars: The Viking Mars Lander. *Icarus.* 1972;16(1):111-138.
67. Boynton WV, Ming DW, Kounaves SP, et al. Evidence for Calcium Carbonate at the Mars Phoenix Landing Site. *Science.* 2009;325(5936):61-64.
68. Figg D, Kahr MS. Elemental fractionation of glass using laser ablation inductively-coupled plasma mass spectrometry. *Applied Spectroscopy.* 1997;51(8):1185-1192.
69. Fryer BJ, Jackson SE, Longerich HP. The design, operation and role of the laser-ablation microprobe coupled with an inductively-coupled plasma mass spectrometer (LAM-ICP-MS) in the Earth sciences. *Canadian Mineralogist.* 1995;33:303-312.
70. Guillong M, Horn I, Gunther D. A comparison of 266 nm, 213 nm and 193 nm produced from a single solid state Nd:YAG laser for laser ablation ICP-MS. *Journal of Analytical Atomic Spectrometry.* 2003;18(10):1224-1230.
71. Jeffries TE, Jackson SE, Longerich HP. Application of a frequency quintupled Nd:YAG source (213 nm) for laser ablation inductively-coupled plasma mass spectrometric analysis of minerals. *Journal of Analytical Atomic Spectrometry.* 1998;13:935–940.
72. Jeffries TE, Perkins WT, Pearce NJG. Comparisons of infrared and ultraviolet laser probe microanalysis inductively-coupled plasma mass spectrometry in mineral analysis. *Analyst.* 1995;120(5):1365-1371.
73. Eggins SM, Kinsley LPJ, Shelley JMG. Deposition and element fractionation processes during atmospheric pressure laser sampling for analysis by ICP-MS. *Applied Surface Science.* 1998;127-129:278-286.
74. Lin Y, Yu Q, Hang W, Huang B. Progress of laser ionization mass spectrometry for elemental analysis — A review of the past decade. *Spectrochimica Acta Part B: Atomic Spectroscopy.* 2010;65(11):871-883.
75. Tang M, Arevalo R, Goreva Y, McDonough WF. Elemental fractionation during condensation of plasma plumes generated by laser ablation: a ToF-SIMS study of condensate blankets. *Journal of Analytical Atomic Spectrometry.* 2015;30(11):2316-2322.
76. Zhang B, He M, Hang W, Huang B. Minimizing Matrix Effect by Femtosecond Laser Ablation and Ionization in Elemental Determination. *Analytical Chemistry.* 2013;85(9):4507-4511.
77. Jenner FE, O'Neill HSC. Major and trace analysis of basaltic glasses by laser-ablation ICP-MS. *Geochemistry, Geophysics, Geosystems.* 2012;13(3).

78. Longerich HP, Günther D, Jackson SE. Elemental fractionation in laser ablation inductively coupled plasma mass spectrometry. *Fresenius' Journal of Analytical Chemistry.* 1996;355(5):538-542.
79. Managadze GG, Wurz P, Sagdeev RZ, et al. Study of the main geochemical characteristics of Phobos' regolith using laser time-of-flight mass spectrometry. *Solar System Research.* 2010;44(5):376-384.
80. Arevalo R, Brinckerhoff W, van Amerom F, et al. Design and demonstration of the Mars Organic Molecule Analyzer (MOMA) on the ExoMars 2018 rover. Paper presented at: 2015 IEEE Aerospace Conference; 7-14 March 2015, 2015; Big Sky, MT, USA.
81. Goesmann F, Brinckerhoff WB, Raulin F, et al. The Mars Organic Molecule Analyzer (MOMA) Instrument: Characterization of Organic Material in Martian Sediments. *Astrobiology.* 2017;17(6-7):655-685.
82. Goetz W, Brinckerhoff WB, Arevalo R, et al. MOMA: the challenge to search for organics and biosignatures on Mars. *International Journal of Astrobiology.* 2016;15(3):239-250.
83. Getty SA, Brinckerhoff WB, Cornish T, Ecelberger S, Floyd M. Compact two-step laser time-of-flight mass spectrometer for in situ analyses of aromatic organics on planetary missions. *Rapid Communications in Mass Spectrometry.* 2012;26(23):2786-2790.
84. Moreno-García P, Grimaudo V, Riedo A, Tulej M, Wurz P, Broekmann P. Towards matrix-free femtosecond-laser desorption mass spectrometry for in situ space research. *Rapid Communications in Mass Spectrometry.* 2016;30(8):1031-1036.
85. Tulej M, Riedo A, Neuland MB, et al. CAMAM: A Miniature Laser Ablation Ionisation Mass Spectrometer and Microscope-Camera System for In Situ Investigation of the Composition and Morphology of Extraterrestrial Materials. *Geostandards and Geoanalytical Research.* 2014;38(4):441-466.
86. Hall DT, Strobel DF, Feldman PD, McGrath MA, Weaver HA. Detection of an oxygen atmosphere on Jupiter's moon Europa. *Nature.* 1995;373:677.
87. Hansen CJ, Esposito L, Stewart AIF, et al. Enceladus' Water Vapor Plume. *Science.* 2006;311(5766):1422-1425.
88. Broadfoot AL, BELTON MJS, TAKACS PZ, et al. Extreme Ultraviolet Observations from Voyager 1 Encounter with Jupiter. *Science.* 1979;204(4396):979-982.
89. Klingelhöfer G, Morris RV, Bernhardt B, et al. Jarosite and Hematite at Meridiani Planum from Opportunity's Mössbauer Spectrometer. *Science.* 2004;306(5702):1740-1745.
90. Baron D, Palmer CD. Solubility of jarosite at 4–35 °C. *Geochimica et Cosmochimica Acta.* 1996;60(2):185-195.
91. Ehlmann BL, Mustard JF, Swayze GA, et al. Identification of hydrated silicate minerals on Mars using MRO-CRISM: Geologic context near Nili Fossae and implications for aqueous alteration. *Journal of Geophysical Research: Planets.* 2009;114(E2):n/a-n/a.
92. Farrand WH, Glotch TD, Rice Jr JW, Hurowitz JA, Swayze GA. Discovery of jarosite within the Mawrth Vallis region of Mars: Implications for the geologic history of the region. *Icarus.* 2009;204(2):478-488.
93. Brown ME, Hand KP. Salts and Radiation Products on the Surface of Europa. *The Astronomical Journal.* 2013;145(4):110.
94. Christner BC, Royston-Bishop G, Foreman CM, et al. Limnological conditions in Subglacial Lake Vostok, Antarctica. *Limnology and Oceanography.* 2006;51(6):2485-2501.

95. Hand KP, Murray AE, Garvin JB, et al. Report of the Europa Lander Science Definition Team. In. Washington, DC: NASA; 2017:264 pp.
96. Driscoll RL, Leinz RW. Methods for Synthesis of Some Jarosites: Techniques and Methods 5-D1. In: Interior UDot, ed. Reston, VA: US Geological Survey; 2005:5 pp.
97. Zubarev RA, Makarov A. Orbitrap Mass Spectrometry. *Analytical Chemistry.* 2013;85(11):5288-5296.
98. Makarov A, Denisov E, Lange O, Horning S. Dynamic Range of Mass Accuracy in LTQ Orbitrap Hybrid Mass Spectrometer. *Journal of the American Society for Mass Spectrometry.* 2006;17(7):977-982.
99. Hashir MA, Stecher G, Mayr S, Bonn GK. Identification of amino acids by material enhanced laser desorption/ionisation mass spectrometry (MELDI-MS) in positive- and negative-ion mode. *International Journal of Mass Spectrometry.* 2009;279(1):15-24.
100. Nishikaze T, Takayama M. Cooperative effect of factors governing molecular ion yields in desorption/ionization mass spectrometry. *Rapid Communications in Mass Spectrometry.* 2006;20(3):376-382.
101. Nishikaze T, Takayama M. Study of factors governing negative molecular ion yields of amino acid and peptide in FAB, MALDI and ESI mass spectrometry. *International Journal of Mass Spectrometry.* 2007;268(1):47-59.
102. Nitta S, Kawasaki H, Suganuma T, Shigeri Y, Arakawa R. Desorption/Ionization Efficiency of Common Amino Acids in Surface-Assisted Laser Desorption/Ionization Mass Spectrometry (SALDI-MS) with Nanostructured Platinum. *The Journal of Physical Chemistry C.* 2013;117(1):238-245.
103. Sanders JD, Grinfeld D, Aizikov K, Makarov A, Holden DD, Brodbelt JS. Determination of Collision Cross-Sections of Protein Ions in an Orbitrap Mass Analyzer. *Analytical Chemistry.* 2018;90(9):5896-5902.
104. Cui J, Yelle RV, Vuitton V, et al. Analysis of Titan's neutral upper atmosphere from Cassini Ion Neutral Mass Spectrometer measurements. *Icarus.* 2009;200(2):581-615.

Table I. Mass defects of peaks derived from jarosite and the In sample stub using single point and linear regression calibration techniques.

| Calibration Technique | Measured Defect Relative to True Mass (in ppm) | | | | | | | | | | | |
|---|---|---|---|---|---|---|---|---|---|---|---|---|
| | $^{16}$O | $^{32}$S | $^{33}$S | $^{34}$S | $^{39}$K | $^{41}$K | $^{54}$Fe | $^{56}$Fe | $^{57}$Fe | $^{113}$In | $^{115}$In |
| $^{115}$In (single point) | -14 | -5.3 | -4.8 | -4.6 | -4.7 | -4.3 | -2.4 | -3.6 | -2.4 | -0.2 | 0.0 |
| $^{113}$In (single point) | -14 | -5.1 | -4.6 | -4.5 | -4.6 | -4.1 | -2.2 | -3.4 | -2.2 | 0.0 | 0.2 |
| $^{113}$In/$^{115}$In (regression) | -24 | -15 | -15 | -15 | -15 | -14 | -12 | -14 | -12 | -10 | -10 |
| *$^{54}$Fe (single point)* | *-12* | *-2.9* | *-2.4* | *-2.2* | *-2.3* | *-1.9* | *0.0* | *-1.2* | *0.0* | *2.2* | *2.4* |
| $^{56}$Fe (single point) | -9.0 | -1.7 | -1.2 | -1.1 | -1.2 | -0.7 | 1.2 | 0.0 | 1.2 | 3.4 | 3.6 |
| *$^{57}$Fe (single point)* | *-12* | *-2.9* | *-2.4* | *-2.2* | *-2.3* | *-1.9* | *0.0* | *-1.2* | *0.0* | *2.2* | *2.4* |
| $^{54}$Fe/$^{56}$Fe/$^{57}$Fe (regression) | -6.9 | 2.1 | 2.6 | 2.7 | 2.6 | 3.1 | 5.0 | 3.8 | 4.9 | 7.2 | 7.4 |
| $^{34}$S (single point) | -9.7 | -0.6 | -0.2 | 0.0 | -0.1 | 0.3 | 2.2 | 1.1 | 2.2 | 4.5 | 4.6 |
| $^{33}$S (single point) | -9.5 | -0.5 | 0.0 | 0.2 | 0.1 | 0.5 | 2.4 | 1.2 | 2.4 | 4.6 | 4.8 |
| $^{32}$S (single point) | -9.0 | 0.0 | 0.5 | 0.6 | 0.5 | 1.0 | 2.9 | 1.7 | 2.9 | 5.1 | 5.3 |
| $^{32}$S/$^{33}$S/$^{34}$S (regression) | -20 | -11 | -10 | -10 | -10 | -10 | -8.1 | -9.2 | -8.1 | -5.8 | -5.7 |

Table II. Isotopic abundances (uncorrected) measured from nine consecutive analyses of synthetic jarosite with increasing laser energy.

| Laser Energy | $^{115}$In | $^{113}$In | $^{57}$Fe | $^{56}$Fe | $^{54}$Fe | $^{41}$K | $^{39}$K | $^{34}$S | $^{33}$S | $^{32}$S |
|---|---|---|---|---|---|---|---|---|---|---|
| 94 µJ | 95.1% | 4.9% | 2.2% | 92.1% | 5.7% | 6.5% | 93.5% | 4.1% | 0.9% | 95.9% |
| 100 µJ (A) | 95.7% | 4.3% | 2.4% | 91.5% | 6.1% | 8.0% | 92.0% | 5.0% | bdl | 95.0% |
| 100 µJ (B) | 95.7% | 4.3% | 2.8% | 90.3% | 7.0% | 8.0% | 92.0% | 5.5% | 0.9% | 94.5% |
| 103 µJ | 95.1% | 4.9% | 2.0% | 91.4% | 6.6% | 6.4% | 93.6% | 4.3% | 0.7% | 95.7% |
| 113 µJ | 95.1% | 4.9% | 2.7% | 90.6% | 6.6% | 7.4% | 92.6% | 5.5% | 0.9% | 94.5% |
| 116 µJ | 96.0% | 4.0% | 2.9% | 90.0% | 7.2% | 8.5% | 91.5% | 5.6% | 1.0% | 94.4% |
| 118 µJ | 95.6% | 4.4% | 2.8% | 90.3% | 6.9% | 8.0% | 92.0% | 5.0% | 0.9% | 95.0% |
| 192 µJ | 95.6% | 4.4% | 2.6% | 90.6% | 6.8% | 7.7% | 92.3% | 5.8% | 1.0% | 94.2% |
| 195 µJ | 96.1% | 3.9% | 2.6% | 90.0% | 7.4% | 8.3% | 91.7% | 5.5% | 0.9% | 94.5% |
| | | | | | | | | | | |
| EXPECTED* | 95.7% | 4.3% | 2.1% | 92.0% | 5.9% | 6.7% | 93.3% | 4.3% | 0.8% | 95.8% |
| Avg Signal | 95.6% | 4.4% | 2.5% | 90.9% | 6.6% | 7.7% | 92.3% | 5.1% | 0.9% | 94.9% |
| *2s* | *0.7%* | *0.7%* | *0.6%* | *1.6%* | *1.1%* | *1.5%* | *1.5%* | *1.2%* | *0.2%* | *1.2%* |
| 2s$_m$ | 0.1% | 0.1% | 0.1% | 0.2% | 0.1% | 0.2% | 0.2% | 0.1% | 0.0% | 0.1% |
| Deviation | -0.2% | 0.2% | 0.4% | -1.2% | 0.7% | 0.9% | -0.9% | 0.9% | 0.1% | -0.9% |

* Expected compositions reflect representative terrestrial abundances per the Commission on Isotopic Abundances and Atomic Weights (CIAAW)

Note: $^{58}$Fe, $^{40}$K, and $^{36}$S were not measured quantitatively (see main text for discussion)

bdl: below detection limit (i.e., SNR <3)